**Increase in the magnitude of the energy barrier distribution in Ni nanoparticles due to dipolar interactions**

S. H. Masunaga,[1,*] R. F. Jardim,[1] R. S. Freitas,[1] and J. Rivas[2]
[1)] *Instituto de Física, Universidade de São Paulo, CP 66318, 05315-970, São Paulo, SP, Brazil.*
[2)] *Departamento de Física Aplicada, Universidade de Santiago de Compostela, Campus Universitario, 15706 Santiago de Compostela, Spain.*

The energy barrier distribution $E_b$ of five samples with different concentrations $x$ of Ni nanoparticles using scaling plots from *ac* magnetic susceptibility data has been determined. The scaling of the imaginary part of the susceptibility $\chi''(\nu, T)$ vs. $T\ln(t/\tau_0)$ remains valid for all samples, which display Ni nanoparticles with similar shape and size. The mean value $\langle E_b \rangle$ increases appreciably with increasing $x$, or more appropriately with increasing dipolar interactions between Ni nanoparticles. We argue that such an increase in $\langle E_b \rangle$ constitutes a powerful tool for quality control in magnetic recording media technology where the dipolar interaction plays an important role.

The dipolar interactions (DIs) in assemblies of magnetic nanoparticles (NPs) constitute a challenging issue in many areas of physics and have triggered renewed interest due to interesting phenomena that emerge from the collective behavior of the NPs.[1] The complexity of this problem originates from the long-range and anisotropic features of DIs.[2] Understanding and controlling the effect of DIs is of paramount importance to the modern technology of magnetic recording media, because the dipolar interaction is responsible for the coupling among the magnetic NPs. For magnetic recording applications, each magnetic NP is treated as an independent magnetic bit. Therefore, it is desirable to estimate and reduce the dipolar interactions of the magnetic assembly. On the other hand, for magnetic logic devices, the goal is to enhance and tailoring the effect related to DIs, because the magnetostatic coupling of ordered magnetic NPs is used to transfer a magnetic bit between two distant points in the array.[2]

Within this context, the frequency dependence of *ac* magnetic susceptibility ($\chi_{ac}$) allows the determination of important parameters of NP assemblies such as the energy barrier ($E_b$) and the attempt time ($\tau_0$) related to the reorientation of the particle magnetic moments. The very small *ac* magnetic field applied in these temperature-dependent measurements results in a slightly modification in the energy barrier and the dynamics of the system is mainly governed by thermally activated processes with intrinsic energy barriers.[3,4] For uniaxial anisotropy, the associated energy barrier is given by $E_b = KV$, where $K$ is the anisotropy constant and $V$ is the volume of the nanoparticles. However, in actual magnetic systems there is always a distribution of particle size or equivalently a volume distribution $[f(V)]$.

The energy barrier distribution $[f(E_b)]$ of a noninteracting NPs assembly, with a given volume distribution and a random distribution of axes-orientations, may be extracted from the imaginary ($\chi''$) component of $\chi_{ac}$ through the relationship[3]

$$f[E_b(\nu,T)] \approx \frac{6K}{\pi M_S^2} \frac{1}{E_b(\nu,T)} \chi''(\nu,T), \quad (1)$$

where $E_b/k_B = T\ln(t/\tau_0)$, $\nu$ is the frequency, $T$ is the temperature, $K$ is the anisotropy constant, $M_S$ is the saturation magnetization, $t$ is the experimental time window of the experiment, and $\tau_0$ is the attempt time.

The $T\ln(t/\tau_0)$ scaling plots of $\chi_{ac}$ have been widely used for characterizing magnetically different NPs systems.[5-9] Within this context, we present here $[f(E_b)]$ distributions, determined by using $T\ln(t/\tau_0)$ scaling plots, of a set of five samples comprised of Ni NPs with different concentrations $x$: 1.9, 2.7, 4.0, 7.9, and 12.8 wt % Ni in order to monitor the effect of increasing DI along the series. Therefore, limiting conditions for applying such a scaling law in systems with weak DIs are provided. The mean energy barrier values ($\langle E_b \rangle$), extracted from the $T\ln(t/\tau_0)$ scaling plots, were found to increase progressively along the series, exhibiting a clear deviation from the behavior of the intrinsic energy barrier associated with the structural parameters of the nanoparticles.

Ni NPs embedded in an amorphous $SiO_2/C$ matrix, with concentrations of 1.9, 2.7, 4.0, 7.9, and 12.8 wt % Ni, were prepared through a modified sol-gel method. Initially, citric acid was dissolved in ethanol and tetraethyl orthosilicate $Si(OC_2H_5)_4$, 98 %]. Nickel (II) nitrate hexahydrate $[Ni(NO_3)_2 \cdot 6H_2O$, 99.9985 %] was added to the latter mixture and mixed for homogenization. Ethylene glycol was thus added to promote the polymerizing reaction when the solution is heated up to 80 °C. The resulted solid resin was heat treated at 300 °C for 3 h, ground in ball mill, and then pyrolyzed in $N_2$ atmosphere at 500 °C. Further details of the method employed for producing the samples are described elsewhere.[10,11]

Measurements of the frequency dependence of $\chi_{ac}(\nu, T) = \chi' + i\chi''$ were performed in a superconducting quantum interference device SQUID magnetometer and in a Physical Property Measurement System PPMS under zero dc magnetic field, an excitation field of 1 Oe, and driving frequencies ($\nu$) varying over five decades (0.033 - 9999 Hz). X-ray powder-diffraction (XRD) patterns were taken using a D8 Advance Bruker-ASX diffractometer with Cu $K_\alpha$ radiation. The XRD analysis indicated that all samples are comprised of crystalline fcc Ni and an amorphous phase related to $SiO_2/C$. The size distributions of the Ni NPs were determined from transmission electron microscopy (TEM) images, which were obtained on two microscopes: (1) JEOL JEM-2010; and (2) Philips CM-12.

Some physical parameters of the series of Ni NPs studied here are summarized in Table I.[12] They indicate that the five samples consist of Ni NPs with a nearly spherical shape, similar diameter of ~ 5 nm, and width of particle volume distribution ($\sigma_V$), which were well described by log-normal functions.[12] Therefore, we have produced a set of samples with very similar size distributions, but with different concentrations $x$ of the magnetic material (Ni). Accordingly, the average distance between Ni NPs decreases with increasing $x$ as reported in Ref. 12.

Table I - Median diameter $d_0$ and width $\sigma_V$ extracted from TEM analysis in Ref. 12. Results from $\chi_{ac}$ analysis of present work are also listed: $\tau_0$ values used for collapsing $\chi''$ vs. $T\ln(t/\tau_0)$ curves and fitted median $E_{b0}$, width $\sigma_{Eb}$, and mean $\langle E_b \rangle$ parameters of $f(E_b)$ function (see Fig. 2).

| TEM | | | $\chi_{ac}$ | | | |
|---|---|---|---|---|---|---|
| $x$ | $d_0$(nm) | $\sigma_V$ | $\tau_0$(s) | $E_{b0}/k_B$(K) | $\sigma_{Eb}$ | $\langle E_{b0}/k_B \rangle$(K) |
| 1.9 | 4.3 | 0.69 | $10^{-9}$ | 166 | 0.71 | 213 |
| 2.7 | 5.0 | 0.60 | $10^{-9}$ | 202 | 0.69 | 256 |
| 4.0 | 4.9 | 0.72 | $10^{-9}$ | 215 | 0.70 | 275 |
| 7.9 | 5.3 | 0.66 | $10^{-9}$ | 307 | 0.70 | 391 |
| 12.8 | 5.5 | 0.78 | $10^{-11}$ | 613 | 0.60 | 735 |

To determine the energy barrier distribution $f(E_b)$ given by Eq. 1, we have followed a specific protocol. Initially, we choose the $\tau_0$ value that superimpose all $\chi''$ vs. $T\ln(t/\tau_0)$ data to a universal curve for different $\nu$ ($t = 1/\nu$) values, as clearly seen in Fig. 1 (not shown for the sample with 2.7 wt % Ni). For systems of noninteracting magnetic NPs or those with very weak dipolar interactions, the data collapse is expected to occur in a wide range of frequency. The best data collapse was found by using $\tau_0 = 10^{-9}$ s and $10^{-11}$ s for the samples with $x \leq 7.9$ and 12.8 wt % Ni, respectively, as listed in Table I. Hence, these universal plots have provided the $\tau_0$ values necessary for calculating $f(E_b)$.

The values of $\tau_0 = 10^{-9}$ s and $10^{-11}$ s obtained in our Ni NPs samples are in excellent agreement with the ones theoretically predicted for ferromagnetic NPs (between $10^{-11}$ and $10^{-9}$ s).[4] These results also indicate that the DI between Ni NPs in our samples is very weak, a feature mirrored in the extracted $\tau_0$ values which are expected to be reduced when DI between NPs is important. The excellent data collapse also suggests that interparticle DI seems to be negligible or very weak in all samples. Although the $\chi''$ vs $T\ln(t/\tau_0)$ analysis indicates negligible dipolar interaction along the series, the extracted $f(E_b)$ distributions may provide additional information, as discussed below.

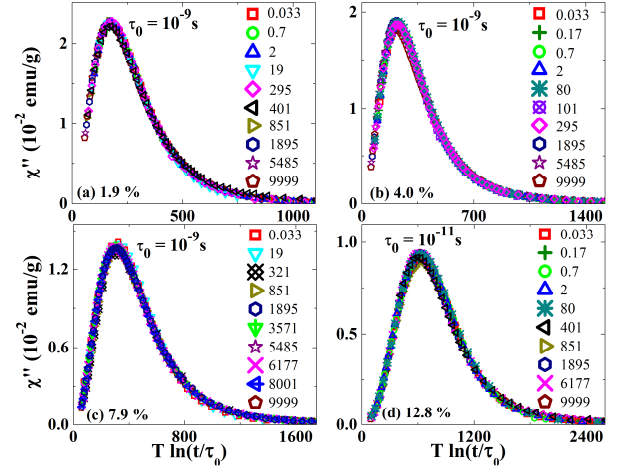

Fig. 1. $\chi''$ vs $T\ln(t/\tau_0)$ scaling plots for the samples with (a) 1.9 %, (b) 4.0 %, (c) 7.9 %, and 12.8 wt % Ni for some selected frequencies.

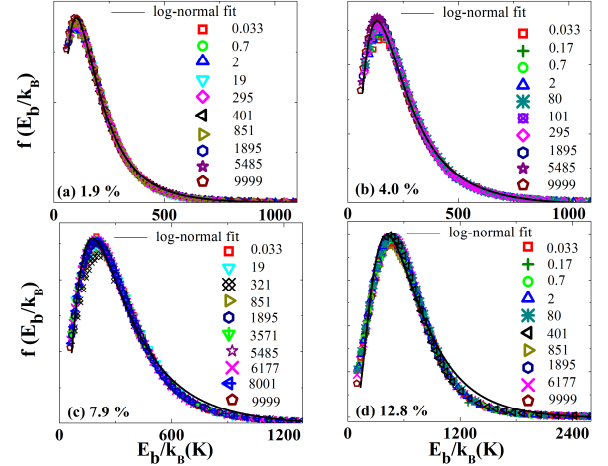

Fig. 2. $f(E_b/k_B)$ given by Eq. 1 for the samples with (a) 1.9 %, (b) 4.0 %, (c) 7.9 %, and 12.8 wt % Ni for some selected frequencies.

The $f(E_b)$ distributions, obtained from Eq. 1, are displayed in Fig. 2. The fitting procedure performed in these curves indicates the same log-normal distribution of particle size, as obtained from TEM analysis. The width of the energy barrier distribution ($\sigma_{Eb}$) is quite similar to the $\sigma_V$ extracted from TEM, as displayed in Table I. However, $E_b$ appears to be Ni concentration dependent (see Table I) and a systematic increase of the mean $\langle E_b \rangle$ value with $x$ is observed. ($E_{b0}/k_B$) varies from 213 K, for the sample with 1.9 wt % Ni, to 735 K for the sample with 12.8 wt % Ni, the latter having the largest Ni concentration of the series. Such an increase in $E_b$ with $x$ deserves additional consideration.

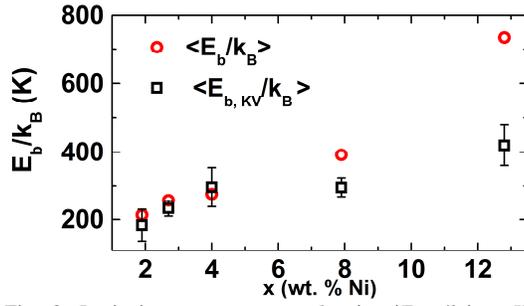

Fig. 3. Intrinsic average energy barrier $\langle E_{b,KV}/k_B \rangle = K_{eff}\langle V \rangle$ and mean $E_b$ ($\langle E_b \rangle = E_{b0} \exp(\sigma_{Eb}^2/2)$) values determined from $f(E_b)$ versus Ni concentration $x$ for the series.

Contributions to $E_b$ may originate from intrinsic anisotropies of the NPs or magnetic interaction. As far as the intrinsic anisotropy is concerned, we have determined the values of $K_{eff}$ to be $4.8 \times 10^5$, $4.1 \times 10^5$, $5.1 \times 10^5$, $4.2 \times 10^5$, and $4.9 \times 10^5$ erg/cm$^3$ for samples with 1.9, 2.7, 4.0, 7.9, and 12.8 wt % Ni, respectively.[12] By using the mean value $\langle V \rangle$, determined from TEM, values of the intrinsic average energy barrier ($\langle E_{b,KV}/k_B \rangle = K_{eff} \langle V \rangle$), in the absence of DI between Ni NPs along the series, were calculated and are displayed in Fig. 3 together with the mean $\langle E_b \rangle$ value determined from $f(E_b)$. The results indicate that $\langle E_{b,KV} \rangle$ is Ni concentration dependent, a feature related to the small difference in the calculated anisotropy constant and average diameter values. In addition to this, values of $\langle E_{b,KV} \rangle$ and $\langle E_b \rangle$ assume almost the same magnitude for the less concentrated samples ($x \leq 4.0$ wt % Ni). However, it is important to notice that the Ni dependence of both $\langle E_{b,KV} \rangle$ and $\langle E_b \rangle$ are quite different for the most concentrated specimens. Such a difference between them increases considerably with increasing $x$ for the most concentrated samples ($x > 4.0$ wt % Ni). The increase in the magnitude of the mean energy barrier, for the two most concentrated samples (7.9 and 12.8 wt % Ni), can not related to the small difference in the volume of the NPs, but instead to the progressive increase of the magnetic interaction along the series. Such an interaction acts in the NPs system increasing their energy barrier in ~ 35 % and ~ 75 % in samples with 7.9 and 12.8 wt % Ni, respectively, when compared to the intrinsic $\langle E_{b,KV} \rangle$. As the Ni NPs are well isolated from each other by the SiO$_2$/C matrix,[12] the increase in the magnitude of the mean $\langle E_b \rangle$ with increasing $x$ is only related to the additional energy barrier created by the DI between Ni NPs.

From the results presented above it follows that the energy barrier distribution and the particle volume distribution are consistent with each other, at least for the most diluted samples (1.9, 2.7, and 4.0 wt % Ni), or more appropriately in samples with negligible and/or very weak DI.[12] The mean $\langle E_b \rangle$ and the $\sigma_{Eb}$ values were found to be in excellent agreement with the ones expected for the intrinsic $\langle E_{b,KV} \rangle$ and $\sigma_V$. For the most concentrated samples (7.9 and 12.8 wt % Ni) we have also obtained excellent scalings of $\chi''(\nu, T)$ data, $\tau_0$ values in the range of the expected ones, and similar results for both $\sigma_{Eb}$ and $\sigma_V$. Nevertheless, the mean $\langle E_b \rangle$ was much larger than expected for the intrinsic $\langle E_{b,KV} \rangle$. Therefore, it is important to remark that, despite the excellent $\chi''(\nu, T)$ vs. $T\ln(t/\tau_0)$ data collapse and $\tau_0$ values extracted from the fitting procedure, the DIs between NPs can not be considered negligible along the series, at least for Ni concentrations greater than about 6 wt % Ni.

The results described here for a set of five samples where the concentration of Ni NPs is the only relevant variable parameter have allowed us to study the effect of the dipolar interactions on the energy barrier distribution of the NPs. The increase of the dipolar interaction between Ni NPs results in an increase in the energy barrier due to dipolar interaction between NPs, despite the excellent $\chi''(\nu, T)$ vs. $T\ln(t/\tau_0)$ data collapse found. Such a feature is accompanied by the agreement of the values obtained for the attempt times $\tau_0$ with theoretical expectations, as discussed for the most concentrated samples with 7.9 and 12.8 wt % Ni. We argue that the combined use of structural data and scaling plots of $\chi''(\nu, T)$ described here provides a sensible protocol to detect small dipolar interactions in assemblies of magnetic nanoparticles. Such a protocol may be used for quality control in devices comprising of magnetic NPs in magnetic recording media where the role of dipolar interaction is crucial for this technology.

The authors acknowledge financial support from the Brazilian agencies FAPESP, CNPq, and CAPES, and Xunta de Galicia and Ministerio de Ciencia e Innovación, Spain.

* suelih@if.usp.br